\documentclass[useASM]{mn2e}

\usepackage{graphicx}
\begin{document}

\title[Radiative efficiency of GRB 070110]
{The radiative efficiency of relativistic jet and wind: A case
study of GRB 070110}

\author[Du et al.]
{Shuang Du$^{1,2}$, Hou-Jun L\"{u}$^{1,2}$\thanks{E-mail: lhj@gxu.edu.cn}, Shu-Qing Zhong$^{1,2}$, En-Wei Liang$^{1,2,3}$\\
 $^1$GXU-NAOC Center for Astrophysics and Space Sciences, Department of Physics, Guangxi University, Nanning 530004,
 China\\
 $^2$Guangxi Key Laboratory for Relativistic Astrophysics, Nanning, Guangxi 530004, China\\
 $^3$National Astronomical Observatories, Chinese Academy of Sciences, Beijing 100012, China\\}
 \maketitle

\label{firstpage}
\begin{abstract}
A rapidly spinning, strongly magnetized neutron star is
invoked as the central engine for some Gamma-ray bursts (GRBs),
especially, the $``$internal plateau$"$ feature of X-ray
afterglow. However, for these $``$internal plateau$"$ GRBs, how
to produce their prompt emission remains an open question. Two
different physical process have been proposed in the
literature, (1) a new-born neutron star is surrounded by a
hyper-accreting and neutrino cooling disk, the GRB jet can be
powered by neutrino annihilation aligning the spin axis; (2) a
differentially rotating millisecond pulsar was formed due to
different angular velocity between the interior core and outer
shell parts of the neutron star, which can power an episodic
GRB jet. In this paper, by analyzing the data of one peculiar
GRB 070110 (with internal plateau), we try to test which model
being favored. By deriving the physical parameters of magnetar
with observational data, the parameter regime for initial
period ($P_{0\rm }$) and surface polar cap magnetic field
($B_{\rm p}$) of the central NS are $(0.96\sim 1.2 )~\rm ms$
and $(2.4\sim 3.7)\times 10^{14}~\rm G$, respectively. The
radiative efficiency of prompt emission is about $\eta_{\gamma}
\sim 6\%$. However, the radiative efficiency of internal
plateau ($\eta_{\rm X}$) is larger than $31\%$ assuming the
$M_{\rm NS}\sim1.4 M_{\odot}$ and $P_{0\rm }\sim1.2 ~\rm ms$.
The clear difference between the radiation efficiencies of
prompt emission and internal plateau implies that they maybe
originated from different components (e.g. prompt emission from
the relativistic jet powered by neutrino annihilation, while
the internal plateau from the magnetic outflow wind).

\end{abstract}

\section {Introduction}
Gamma-ray bursts (GRBs) are the most luminous events ever known
in the universe by far. Traditionally, the relativistic
fireball model is proposed to interpret the observational
phenomenon of GRBs (M\'{e}sz\'{a}ros 2002; Zhang \&
M\'{e}sz\'{a}ros 2004; Kumar \& Zhang 2015). Within this
scenario, the observed prompt gamma-ray emission is explained
by the internal shocks (M\'{e}sz\'{a}ros \& Rees 1993). Despite
its attractive features, the internal shock model is suffered
with some severe problems, such as inefficiency problem (Kumar
\& Zhang 2015 for a review). Alternatively, if the outflow is
dominated by Poynting flux, significant magnetic energy is
dissipated to produce non-thermal emission, a fraction of the
dissipated energy is converted to kinetic energy (Zhang \& Yan
2011). After the internal dissipation, the deceleration of the
jet by the ambient medium excites a long term external shock
with synchrotron emission which powers the broad band afterglow
emission (M\'{e}sz\'{a}ros \& Rees 1997; Sari, Piran \& Narayan
1998; Zhang et al. 2006; Gao et al. 2013). In {\em Swift} era,
the shallow decay (or plateau) segment is usually seen in the
XRT light curves (Liang et al. 2007), and the widely discussed
model for this component is energy injection into the external
forward shock either from an long lasting central engine or
from an ejecta with a wide distribution of Lorentz factors
(Zhang et al. 2006; Nousek et al. 2006; Panaitescu et al.
2006). On the other hand, in rare cases, X-ray plateaus of long
GRBs can be followed by a very steep decay (e.g. $t^{-9}$; GRB
070110, Troja et al. 2007; L\"{u} \& Zhang
2014)\footnote{Throughout the paper, we use the convention
$f\propto t^{-\alpha}\nu^{-\beta}$ for temporal and spectral
power law models.}, and in some short GRBs as well (Rowlinson
et al. 2010, 2013; L\"{u} et al. 2015), which called an
$``$internal plateau$"$ (Lyons et al. 2010). This is more
difficult to be explained by standard external afterglow
fireball model, and an internal dissipation process need to be
invoked (Fan \& Xu 2006). From theoretical point of view, such
behavior could be naturally explained when a rapidly spinning,
strongly magnetized neutron star called $``$millisecond
magnetar$"$ being invoked as the central engine of GRB (Dai \&
Lu 1998a,b; Zhang \& M\'{e}sz\'{a}ros 2001; Gao \& Fan 2006;
Metzger et al. 2008), and the internal plateau feature is also
$``$Smoking Gun$"$ signature for magnetar collapsing into black
hole (Kumar \& Zhang 2015 for review).

Previous work have shown that rapidly spinning, strongly
magnetized NS could produce both prompt emission and later
plateau of afterglow with proper parameters (Usov et al. 1992).
Under this framework, two different physical process are
proposed to produce GRB jet. One is that a new-born neutron
star surrounded by hyper-accreting and neutrino cooling disk,
which is similar to disk cooling of black hole central engine
via neutrino annihilation (Zhang \& Dai 2008, 2009; Lei et al.
2009, 2013), but the structure of a hyper-accretion disk may be
different. Zhang \& Dai (2008, 2009) divide the disk of neutron
star into two regions (e.g. inner and outer disks), and studied
physical properties of disk structure. The GRB hot jet can be
powered by neutrino annihilation following the spin axis.
Later, the magnetar would release its rotation energy via
magnetic dipole radiation to produce the observational plateau
in X-ray afterglow. Alternative, a differentially rotating
millisecond pulsar was formed due to different angular velocity
between the interior core and outer shell parts of the neutron
star. It wind up toroidal magnetic fields to about $10^{15-16}$
G, and release the corresponding magnetic energy via magnetic
reconnection or magnetic dissipation instabilities when each
buoyant magnetic field torus floats up to break through the
stellar surface (Klu\'{z}niak \& Ruderman, 1998; Dai et al.
2006). This released energy can be satisfied with prompt
emission requirement of GRBs observations. After that, magnetar
continuously spin down, residual rotational energy would be
dissipated by magnetic dipole radiation to power the X-ray
internal plateau. Also, similar physical process was proposed
to produce relativistic jet and later X-ray plateau with a
varying $\sigma$ value in different phase (Metzger et al.
2011).

One interesting question is that how to distinguish the origin
of its prompt emission, which is important for understanding
the composition of the jet. Within those two different physical
process, from the theoretical point of view, if the GRB hot jet
powered by neutrino annihilation, it predicts a lower radiative
efficiency, typically a few percent (Kumar 1999; Panaitescu et
al. 1999; Kumar \& Zhang 2015). In contrary, if the magnetic
dissipation is dominated in the prompt emission, it is along
with high efficiency, as high as $90\%$ depending on the
$\sigma$ value (Zhang \& Yan 2011). To distinguish these two
models, we suggest to compare the radiation efficiency between
the prompt emission phase and internal plateau phase. It is
wildly accepted that the internal plateau phase is from
magnetar wind dissipation process, so that if the radiation
efficiency of the prompt phase is similar to the plateau phase,
magnetic dissipation of magnetar is favored, otherwise neutron
star surrounded by a hyper-accreting would be favored.

This paper is to address this interesting question through
analyze the data of GRB 070110. The XRT data reduction and fitting
are presented in \S 2. In \S 3, we show detail calculations of
radiation efficiency. Finally, the conclusions and discussion
are given in \S 4. Throughout, a concordance cosmology with
parameters $H_0 = 71$ km s$^{-1}$ Mpc $^{-1}$, $\Omega_M=0.30$,
and $\Omega_{\Lambda}=0.70$ are adopted.

\section{Multiple wavelength observations and calculations of GRB 070110}

So far, more than 120 GRBs have been observed with shallow (or
plateau) decay segment in the X-ray afterglow. However, if a
normal decay is followed the plateau, it can not be confident
to show that the shallow decay is originated from the internal
dissipation of magnetar spin-down  (Panaitescu et al. 2006). In
order to find out the magnetar signature, which typically
invokes a shallow decay phase (or plateau) followed by a
steeper decay segment (steeper than $t^{-3}$). One requires
three independent criteria to define our sample. First, it
displays an $``$internal plateau$"$. Second, after the sharp
decay following with plateau, another power-law component is
appeared with decay index less than 1.5, which is contributed
by the external shock emission. Third, the redshift of the
burst need to be measured, in order to estimate the gamma-ray
energy and kinetic energy. We systematically process the XRT
data of more than 1250 GRBs observed between 2005 January and
2016 March. Only GRB 070110, with duration $T_{90}\sim 88$s, is
satisfied with those three requirements in our entire sample.
We next perform a temporal fit to the plateau behavior of GRB
070110 with a smooth broken power law
\begin{eqnarray}
F = F_{0} \left[\left(\frac{t}{t_b}\right)^{\omega\alpha_1}+
\left(\frac{t}{t_b}\right)^{\omega\alpha_2}\right]^{-1/\omega},
\end{eqnarray}
add single power-law function
\begin{eqnarray}
F = F_{1}t^{-\alpha_3} ,
\end{eqnarray}
where $t_{b}$ is the break time, $F_b=F_0 \cdot 2^{-1/\omega}$
is the flux at the break time $t_b$,  $\alpha_1$, $\alpha_2$
and $\alpha_3$ are decay indices, respectively, and $\omega$
describes the sharpness of the break. The larger the $\omega$
parameter, the sharper the break. We also collect the optical
observational data from Troja et al (2007). Both X-ray and
optical light curve are shown in Figure 1, and fitting result
is presented in Table 1.

Another two important parameters are the isotropic gamma-ray
energy ($E_{\rm \gamma,iso}$) and kinetic energy ($E_{\rm
K,iso}$). $E_{\rm \gamma,iso}$ was measured from the
observation flunce and distance, read as
\begin{eqnarray}
E_{\rm \gamma,iso}&=&4\pi k D^{2}_{L} S_{\gamma} (1+z)^{-1}\nonumber \\
 &=&(3.09\pm2.51)\times 10^{52}~ {\rm erg}
\end{eqnarray}
where $z=2.352$ is the redshift, $D_{L}$ is the luminosity
distance, $S_{\rm \gamma}=(1.8\pm0.2)\times 10^{-6} \rm erg~
cm^{-2}$ is gamma-ray fluence in BAT band, and $k$ is the
$k$-correction factor from the observed band to $1-10^4$ keV in
the burst rest frame (e.g. Bloom et al. 2001). More details,
please refer to L\"{u} \& Zhang (2014). The $E_{\rm K,iso}$, is
isotropic kinetic energy of the fireball. It could be estimated
by standard forward afterglow model (Sari, Piran \& Narayan
1998; Fan \& Piran 2006). For the late time X-ray afterglow
data ($t> 5\times10^4$s), one has decay slope $\alpha_3\sim
0.82$, and the spectral index $\beta_{\rm X}\sim 1.12$ in the
normal decay segment. Approximately, they are satisfied
$2\alpha_3\simeq 3\beta_{\rm X}-1$ in the spectral regime $\nu
> max(\nu_{\rm m},\nu_{\rm c})$, where $\nu_{\rm c}$ and
$\nu_{\rm m}$ are the typical and cooling frequencies of
synchrotron radiation, respectively. Following the equations
and methods of Yost et al (2003), the flux was recorded in XRT
(0.3 keV - 10 keV) as,
\begin{eqnarray}
\rm Flux &=& 1.2\times10^{-12} ~{\rm erg~s^{-1}~cm^{-2}}(\frac{1+z}{2})^{(p+2)/4}D_{L,28}^{-2} \nonumber  \\& \times &
\epsilon_{B,-2}^{(p-2)/4}\epsilon_{e,-1}^{p-1}E_{\rm K,iso,53}^{(p+2)/4}(1+Y)^{-1}t_{d}^{(2-3p)/4},
\end{eqnarray}
in this calculation, the Compton parameter (Y) is assigned to a
typical value $Y=1$. Combine with the observational data, one
obtain $E_{\rm K, iso}\sim 5\times 10^{53} \rm ~erg$, the
physical parameters of forward shock model are shown in Table
1, and the fitting result is presented in Figure 1.

\section{Prompt emission and Radiative efficiency of GRB 070110}

\subsection{Physical parameters of magnetar for GRB 070110}

Since the internal plateau of GRB 070110 was explained by
invoking magnetic dipole radiation of spin-down magnetar
central engine. In this section, we use data to derive relevant
physical magnetar parameters of GRB 070110 (e.g. the initial
spin period $P_0$ and the surface polar cap magnetic field
$B_p$).

The energy reservoir is the total rotation energy of the
millisecond magnetar, which reads
\begin{equation}
E_{\rm rot} = \frac{1}{2} I \Omega_{0}^{2}
\simeq 2 \times 10^{52}~{\rm erg}~
M_{1.4} R_6^2 P_{0,-3}^{-2},
\label{Erot}
\end{equation}
where $I$ is the moment of inertia, $\Omega_0 = 2\pi/P_0$ is
the initial angular frequency of the neutron star, $M_{1.4} =
M/1.4M_\odot$, $R$ is radius of NS, and the convention $Q =
10^x Q_x$ is adopted in cgs units for all other parameters
throughout the paper. Assuming that the magnetar with initial
spin period $P_0$ is being spun down by a magnetic dipole with
surface polar cap magnetic field $B_p$, the characteristic
spin-down luminosity and spin-down time scale are
\begin{equation}
 L_0 = 1.0 \times 10^{49}~{\rm erg~s^{-1}} (B_{p,15}^2 P_{0,-3}^{-4} R_6^6)
\label{L0}
\end{equation}
\begin{equation}
 \tau = 2.05 \times 10^3~{\rm s}~ (I_{45} B_{p,15}^{-2} P_{0,-3}^2 R_6^{-6})
\label{tau}
\end{equation}
The internal plateau energy of GRB 070110 from internal
dissipation ($E_{\rm pla}$) is calculated based on the
lightcurve fitting result and redshift information, read as
\begin{eqnarray}
E_{\rm X,iso,pla}&=&\int_{t_{s}}^{t_{b}}\frac{L_{\rm pla}}{1+z}dt \nonumber \\
&=&\frac{4\pi D_{L}^{2}}{(1+z)}\int_{t_{s}}^{t_{b}}F_{b}t^{\alpha _{1}}dt \nonumber \\
&=&(4.3\pm0.4)\times 10^{51}~ {\rm erg}
\end{eqnarray}
where $t_s$ and $t_b$ is the starting and end time of internal
plateau. Here, we adopt $t_s=0$. Actually, the starting time is
not effect the result too much because $t_s$ is much less than
$t_b$.

On the other hand, two additional constraints are required to
be satisfied in this situation. Firstly, the spin-down
luminosity of magnetar should be brighter than observational
internal plateau luminosity of GRB 070110 if internal plateau
emission is contributed from magnetic dipole radiation, namely
$L_{0}>L_{\rm pla}$. Another one is that spin-down time scale
is larger than duration of internal plateau (maybe collapse
time of magnetar into black hole), $\tau > t_b$. Use those two
constraints with lower limit of initial period NS survived, the
region of initial period and surface polar cap magnetic field
of NS are $(0.96\sim 1.2 )\rm~ ms$ and $(2.4\sim 3.7)\times
10^{14}\rm~ G$, respectively. The result is shown in Figure 2
(gray region).

\subsection{Radiative efficiency of relativistic jet and wind}
One interesting question is that what is the radiative
efficiency of GRB 070110 for prompt emission and internal
plateau within the GRB jet produced by neutrino annihilation
scenario. The GRB radiation efficiency of prompt emission is
defined as (Lloyd-Ronning \& Zhang 2004)
\begin{eqnarray}
\eta_{\gamma} = \frac{E_{\rm \gamma,iso}}{E_{\rm \gamma,iso}+E_{\rm K,iso}} = \frac{E_{\rm \gamma}}{E_{\rm \gamma}+E_{\rm K}}
\end{eqnarray}
where $E_{\rm \gamma,iso}\sim (3.09\pm 2.51)\times 10^{52} \rm~
erg$ and $E_{\rm K,iso}\sim 5\times 10^{53} \rm~ erg$. One has
$\eta_{\gamma} \sim (6\pm 4)\%$.

Another radiative efficiency is from internal dissipation of
internal plateau, which is defined as the ratio between
internal plateau energy and total magnetic dipole radiation
energy of magnetar ($E_{\rm m}$), read as
\begin{eqnarray}
\eta_{\rm X} = \frac{E_{\rm X,iso,pla}}{E_{\rm m}}.
\end{eqnarray}
It reflects how efficient the internal dissipation converts the
total magnetic dipole energy into radiation during the X-ray
internal plateau phase. The total magnetic dipole radiation
energy $E_{\rm m}$ should be less than $E_{\rm rot}$, namely
$E_{\rm m} < E_{\rm rot}$, one can get the lower limit of
efficiency of internal plateau $\eta_X>31\%$ for $M_{\rm
NS}\sim1.4M_{\odot}$ and $P_0\sim1.2 \rm~ ms$.

The clear difference between the radiation efficiencies of
prompt emission and internal plateau implies that they may be
originated from different components, e.g. prompt emission from
the relativistic jet powered by neutrino annihilation, while
the internal plateau from the magnetic outflow wind.

\section{Conclusions and Discussion}
An internal dissipation process of magnetar with Poynting-flux
dominated outflow was invoked to interpret internal plateau
phase of GRB afterglows. We suggest that comparing the
radiation efficiency of prompt emission and internal plateau
phase could help to investigate the composition of GRB jet. We
focus on analyzing the data of GRB 070110 which exhibits
internal plateau feature following a normal decay. We firstly
estimate the physical parameters of magnetar based on the
observational feature of internal plateau, the parameter regime
of initial period ($P_{\rm 0}$) and surface polar cap magnetic
field ($B_{\rm p}$) of NS are $(0.96\sim 1.2)\rm~ ms$ and
$(2.4\sim 3.7)\times 10^{14}\rm~ G$, respectively. In this
case, the radiation efficiency of prompt emission would be
$\eta_{\gamma} \sim (6\pm 4)\%$ if the GRB jet was powered by
neutrino annihilation. On the other hand, the lower limit of
internal plateau radiative efficiency is estimated as
$\eta_{\rm X}=31\%$ with $M_{\rm NS}\sim1.4 M_{\odot}$ and
$P_{0\rm }\sim1.2 \rm ~ms$.

Since the standard internal shock model and magnetic
dissipation model for prompt emission predict lower and higher
radiation efficiency, respectively (Kumar 1999; Panaitescu et
al. 1999; Usov 1992; Zhang \& Yan 2011). Also, it is wildly
accepted that the internal plateau phase is from magnetar wind
dissipation process, and the prompt emission radiation
efficiency ($\eta_{\gamma} \sim 6\%$) is much less than the
minimum efficiency of internal plateau ($\eta_{\rm X}=31\%$),
so that the prompt emission and later internal plateau of GRB
070110 may be from different origin, e.g., a new-born neutron
star surrounded by a hyper-accreting disk generates the prompt
emission, while the magnetic dipole dissipation is account for
the later internal plateau.

One suspicion is that whether the neutrino annihilation of NS
cooling can power the prompt emission of GRB 070110. If neutron
star surrounded by a hyper-accreting model was accepted to
power the GRB jet, the neutrino annihilation luminosity
($L_{\rm \nu\bar{\nu}}$) is contributed by neutrinos emitted
from both disk and neutron star surface layer. Following Zhang
\& Dai (2009) method, there is no analytical solution of
$L_{\rm \nu\bar{\nu}}$, but related to several parameters, e.g.
accretion rate ($\dot{M}$), outflow index ($s$), viscosity
($\alpha$), energy parameter ($\varepsilon$) and efficiency
factor to measure the surface emission ($\eta_{\rm s}$).
Therefore, we have to use numerical method to get the solution
with right parameters to compare with observational prompt
emission luminosity. Since $L_{\rm \nu\bar{\nu}}$ are not
sensitively depending on the $\alpha$, $\varepsilon$ and $s$
(see Figure 7 and 8 in Zhang \& Dai 2009), we fix the typical
value of $\alpha=0.1$, $\varepsilon=0.5$ and $s=0.2$. Assuming
$\eta_{\rm s}=0.5$ and $\dot{M}=0.03~\rm M_{\odot}~s^{-1}$, one
has $L_{\rm \nu\bar{\nu}}\sim 3\times 10^{48}~\rm erg~s^{-1}$.
However, it is isotropic energy instead of true energy. Due to
lack observation of jet break feature, one can estimate lower
limit of the jet opening angle with the last observed point
($t_{j}\sim 25$ days) in X-ray afterglow, read as
\begin{eqnarray}
\label{theta_j}
\theta_j  &=&  0.057 ~\rm rad ~\left(\frac{t_j}{1 ~\rm day}\right)^{3/8}\left(\frac{1+z}{2}\right)^{-3/8} \nonumber  \\& \times &
\left(\frac{E_{\rm K,iso}}{10^{53}~\rm ergs}\right)^{-1/8}\left(\frac{n}{0.1 ~\rm cm^{-3}}\right)^{-3/8}\nonumber  \\&=&
7.4^{\circ}
\end{eqnarray}
The prompt emission energy of GRB jet after beaming-corrected
is
\begin{eqnarray}
E_{\rm \gamma}=E_{\rm \gamma, iso}\cdot f_{b}\simeq 2.5\times 10^{50} \rm~erg
\end{eqnarray}
where $f_b$ is beaming factor of the GRB 070110
\begin{eqnarray}
\label{fb}
f_b = 1-\cos \theta_j \simeq (1/2) \theta_j^2,
\end{eqnarray}
and the luminosity of prompt emission is $L_{\rm jet}\sim
E_{\rm \gamma}/T_{90}\sim 2.8\times 10^{48} ~\rm erg~s^{-1}$.
One has $L_{\nu\dot{\nu}}> L_{\rm jet}$ with typical value of
parameters and $\dot{M}=0.03~\rm M_{\odot}~s^{-1}$, namely,
neutrino annihilation of NS can provide enough energy to power
the GRB jet.

\section{Acknowledgements}
We acknowledge the use of the public data from the Swift data
archive, and the UK Swift Science Data Center. We thank Wei-Hua
Lei for helpful comments and discussion. This work is supported
by the National Basic Research Program (973 Programme) of China
2014CB845800, the National Natural Science Foundation of China
(Grant No. 11533003), the One-Hundred-Talents Program of
Guangxi colleges, Guangxi Science Foundation (grant No.
2013GXNSFFA019001), Scientific Research Foundation of GuangXi
University (Grant No XGZ150299).

%*******************************************************************************************
\begin{figure}
\center
\includegraphics[angle=0,scale=0.4]{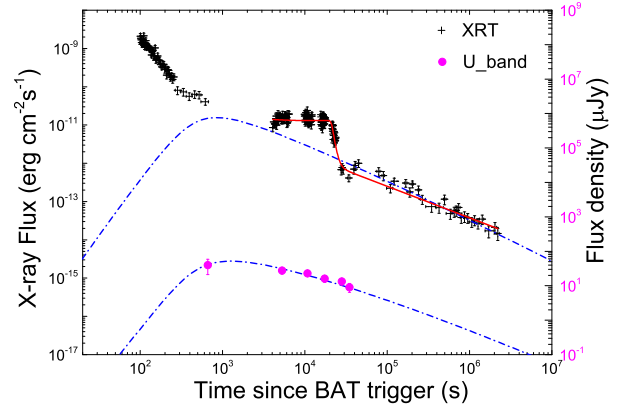}
\caption{The X-ray (black plus) and optical (magenta dots) lightcurve of GRB 070110.
The solid red line is the empirical function fitting. The dash lines are numerical
calculations by standard external forward shock model.}
\end{figure}
%*******************************************************************************************

\begin{figure}
\center
\includegraphics[angle=0,scale=0.4]{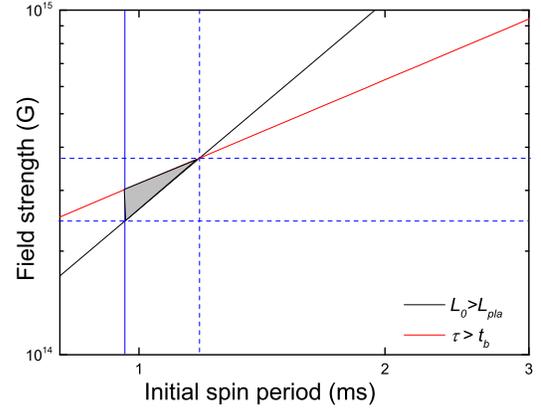}
\caption{Initial spin period $P_0$ {\em vs} surface polar cap magnetic
field strength $B_p$ distributions, which derived from two independent conditions.
The gray region is an available parameters distributions for GRB 070110 magnetar central engine.
The vertical solid line is the breakup spin-period for a neutron star (Lattimer \& Prakash 2004).}
\end{figure}
%*******************************************************************************************

\begin{table}
  \begin{tabular}{cccccccccc}
  \hline
  \multicolumn{2}{c}{Lightcurve fitting} &{}{}&
  \multicolumn{2}{c}{External forward shock}\\
  \hline
  Parameters & Value  & {} & Parameters & Value \\
  \hline
  $F_{0} $& (1.23$\pm$0.06)e-11 $~\rm erg~cm^{-2}~s^{-1}$&  {}& $\epsilon_{e}$& 0.02\\
  $F_{1}$& (3.23$\pm$1.58)e-9 $~\rm erg~cm^{-2}~s^{-1}$ &  {} &$\epsilon_{b}$ & 5e-4\\
  $\alpha_{1}$ & $0.10\pm0.07$ & {}& $n$& $5~\rm cm^{-3}$\\
  $\alpha_{2}$& $8.7\pm0.8$ & {}& $\Gamma$& 90\\
  $\alpha_{3}$& $ -(0.82\pm0.04) $ & {} & $p$& 2.2\\
  $t_{b}$& $20885\pm222 \rm~s$ &  {} &$E_{\rm K,iso}$ & 5e53 $~\rm erg$\\
   \hline
\end{tabular}
\caption{The fitting result of GRB 070110 lightcurve with
smooth broken power-law and single power-law, and the
parameters of standard external forward shock model.}
\end{table}
%**********************************************************************

\end{document}